# Dislocation core properties of β-tin: A first-principles study

M.A. Bhatia[1], M. Azarnoush[1], I. Adlakha[1], G. Lu[2] and K.N. Solanki[1]*

[1]*School for Engineering of Matter, Transport, and Energy; Arizona State University, Tempe, AZ*

[2]*Department of Physics and Astronomy; California State University Northridge, Northridge, CA*

*(480)965-1869; (480)727-9321 (fax), E-mail: kiran.solanki@asu.edu, (Corresponding author)

**Abstract**

Dislocation core properties of tin (β-Sn) were investigated using the semi-discrete variational Peierls-Nabarro model (SVPN). The SVPN model, which connects the continuum elasticity treatment of the long-range strain field around a dislocation with an approximate treatment of the dislocation core, was employed to calculate various core properties, including the core energetics, widths, and Peierls stresses for different dislocation structures. The role of core energetics and properties on dislocation character and subsequent slip behavior in β-Sn was investigated. For instance, this work shows that a widely spread dislocation core on the {110} plane as compared to dislocations on the {100} and {101} planes. Physically, the narrowing or widening of the core will significantly affect the mobility of dislocations as the Peierls stress is exponentially related to the dislocation core width in β-Sn. In general, the Peierls stress for the screw dislocation was found to be orders of magnitude higher than the edge dislocation, i.e., the more the edge component of a mixed dislocation, the greater the dislocation mobility (lower the Peierls stress). The largest Peierls stress observed was 365 MPa for the dislocation on the {101} plane. Furthermore, from the density plot, we see a double peak for the 0° (screw) and 30° dislocations which suggests the dissociation of dislocations along these planes. Thus, for the {101} <$\bar{1}$01> slip system, we observed dislocation dissociation into three partials with metastable states. Overall, this work provides qualitative insights that aid in understanding the plastic deformation in β-Sn.

## I. Introduction

Lead-free solder materials, such as high-tin (Sn) alloys and electrically conductive adhesives (ECAs) have been considered as promising alternatives in electronic packaging as they do not present adverse health risks [1,2]. However, the performance demands placed on advanced electronic packages drives a reduction in solder bump size and pitch. As solder volume decreases there is an increased likelihood for undercooling to occur, and the resulting joint may consist of only a few large grains [3]. Thus, Sn rich alloys are known to fail due to the nucleation and growth of intermetallic phases that drastically alter the mechanical properties of the Sn-alloy [1,4]. Furthermore, at room temperature, β-Sn has highly anisotropic elastic and plastic responses [5–7] due to the body-centered-tetragonal (BCT) crystal structure. Therefore accurate characterization of anisotropic mechanical behavior in Sn grains is increasingly important for solder joint reliability [8]. Recently, Bieler et al. [1,9,10] have analyzed deformation in Sn-Ag solder lap shear joints under thermomechanical fatigue using orientation imaging microscopy (OIM). They found that grain boundary sliding occurred and that the slip behavior could be described using Schmid factor analysis. Conversely, Matin et al. [11] studied the slip behavior in bulk Sn-3.8Ag-0.7Cu fatigue samples and found that fatigue damage was not localized at the grain boundaries; rather, it occurred in the form of persistent slip bands within favorably oriented grains. Indeed, Sidhu and Chawla [12] found that the mechanical constraint imposed on the solder volume dominates over the grain orientation effects in determining the location of crack initiation. Therefore, in order to accurately determine the orientation dependent mechanical behavior of Sn-rich alloys it is clear that understanding all possible active slip systems is necessary along with analyzing the dislocation core properties and glide resistance [2,6,7,13].

Quantum mechanics- and crystal plasticity mechanics-based simulations are increasingly utilized in investigations of fundamental plasticity mechanisms in β-Sn. For instance, based on first-principles calculations, Kinoshita et al. [2] show that the (101) [$\bar{1}$01] slip system had the least ideal shear strength. Similarly, Bhatia et al. [14], using generalized stacking fault energies (GSFE) of different slip systems, concluded that the {100} <001> slip system is the most easily activated system. So far, however, no studies have been conducted to address the sensitivity of slip behavior on the dislocation structure, which plays a crucial role in the plasticity of $\beta$-Sn [15]. Note that there are about 32 different slip systems in BCT tin and we have only a semi-quantitative understanding of their rank and likelihood to contribute to deformation [10]. The determination of the correct slip systems and the elucidation of their deformation mechanisms is a crucial first-step towards the development of discrete dislocation and crystal plasticity models, which can then predict dislocation cell substructure and creep behavior [16]. Therefore, in this work, the preferential slip systems in β-tin were investigated using the first-principles based SVPN method.

In the present paper, the first-principles SVPN model connects the continuum elasticity treatment for the long-range strain field around a dislocation with an approximate treatment of the dislocation core through the first-principles determined GSFE curves [17,18]. This method was employed to calculate various core properties, including the core energetics, widths, and Peierls stresses for different dislocations along with the dissociation of perfect dislocations into partial dislocations in β-Sn at 0 K. Note that in the context of crystal plasticity, the SVPN model provides the upper bound on the dislocation mobility by revealing the nature of core narrowing or broadening in the glide planes. Thus, the role of core energetics and properties on dislocation character and subsequent slip behavior in β-Sn were investigated. Based on the SVPN model, it was concluded that i) {101} <10$\bar{1}$>, with a <c+a> type Burger's vector, is the most easily activated slip system; ii) the {100} <001>, with a <c> type Burger's vector, is the slip system which has the lowest Peierls stress, and thus, easy dislocation glide as compared to other slip systems, and iii) there is a significant anisotropy at the dislocation core in Sn. Further, the Peierls stress for the screw dislocation was found to be orders of magnitude higher than the edge dislocation, i.e., the more the edge component of a mixed dislocation, the greater the dislocation mobility (lower the Peierls stress). The largest Peierls stress observed was 365 MPa for the dislocation on the {101} plane. The simple and computationally expedient nature of the SVPN model makes it easier to provide qualitative insights on the dislocation properties across a large number of slip systems in β-Sn. Furthermore, the SVPN model was able to predict accurate dislocation core structures for fcc metals, such as Al, however; it could overestimate the Peierls stress by a factor of 3 – 4 [19,20]. Thus, in the future, systematic modeling efforts (explicit dislocation core modeling) are necessary in the development of accurate predictive capabilities that can aid in improving the reliability of solder joints. Nevertheless, in this work for the first time a systematic study was carried out to assess the dislocation properties along the various non-equivalent slip system in β-Sn.

## 2. Semi-discrete variational Peierls-Nabarro model

The Peierls-Nabarro (PN) model describes the dislocation core structure within a continuum framework and can be used to estimate the critical stress required for dislocation motion, i.e., Peierls stress [7,13,21–24]. However, the inconsistent treatment of the elastic energy and the misfit energy results in inaccurate predictions of the dislocation core properties, see detail in [7,19,21]. To overcome these shortcomings, the SVPN model was put forward, in which both the elastic and misfit energy are discretized and formulated in a similar manner [7,19,21]. As a result, the total energy of the dislocation becomes variational and the Peierls stress can be determined in a straightforward manner.

The SVPN model has been widely used to predict dislocation core properties in conjunction with first-principles calculations [19,21,24]. In this method, the equilibrium dislocation core structure can be obtained by minimizing the total energy ($U_{dis}$) with respect to the dislocation density (or equivalently the disregistry vector) as a function of applied stress. The total energy ($U_{dis}$) includes four energy contributions:

$$U_{dis} = U_{elastic} + U_{misfit} + U_{stress} + Kb^2 \ln L \tag{1}$$

where *L and b* are the outer cutoff radius for the elastic energy and magnitude of the Burger's vector, respectively. *K* is the pre-logarithmic energy factor,

$$K = \frac{\mu}{2\pi}(\frac{sin^2\theta}{1-v} + cos^2\theta) \tag{2}$$

where $\theta$ is the angle between the Burger's vector and the dislocation line. $\mu$ and $v$ are the shear modulus and Poisson's ratio, respectively. The first term in Eq. (1) is the discretized elastic energy, $U_{elastic}$, and is given by

$$U_{elastic} = \sum_{i,j} \frac{1}{2} \chi_{ij} [K_e(\rho_i^{(1)}\rho_j^{(1)} + \rho_i^{(2)}\rho_j^{(2)}) + K_s\rho_i^{(3)}\rho_j^{(3)}] \tag{3}$$

where the superscripts 1, 2 and 3 correspond to the edge, vertical and screw components of the variables, respectively, and subscript *i* represents the *i*<sup>th</sup> nodal point. The general interplanar dislocation density at the $i^{th}$ nodal point, $\rho_i^{(l)}$, is defined as $\rho_i^{(l)} = (f_i^{(l)} - f_{i-1}^{(l)})/(x_i - x_{i-1})$, where $f_i^{(l)}$ and $x_i$ are the disregistry vector and the coordinate of the $i^{th}$ nodal point (atomic row), respectively. $K_e = \mu/(2\pi(1-\nu))$ and $K_s = \mu/2\pi$ are the energy factors for an edge and a screw dislocation, respectively. The second term in Eq. (1) is the misfit energy, $U_{misfit}$, which is a function of the disregistry vector ($\bar{f}_l$), and is given by

$$U_{misfit} = \sum_i \Delta x \, \gamma_3(\bar{f}_l) \tag{4}$$

where $\gamma_3$ is the three dimensional misfit potential computed using DFT [25,26]. The corresponding components of the applied stress interacting with $\rho_i^{(1)}$, $\rho_i^{(2)}$ and $\rho_i^{(3)}$ are $\tau^{(1)} = \sigma_{21}$, $\tau^{(2)} = \sigma_{22}$ and $\tau^{(3)} = \sigma_{23}$, respectively.

For convenience, the core energy ($U_{core}$) is often defined as the sum of the elastic and misfit energies, which is the dislocation configuration-dependent part of the total energy. The third term in Eq. (1) represents the elastic work done on the dislocation given by

$$U_{stress} = -\sum_{i,l} \frac{x_i^2 - x_{i-1}^2}{2} (\rho_i^{(l)} \tau_i^{(l)}) \tag{5}$$

The last term in Eq. (1), i.e., $Kb^2 \ln L$, is independent of the dislocation core structure. The x-axis is considered the dislocation gliding direction. The rest of the quantities can be computed by the following equations

$$\chi_{ij} = \frac{3}{2}\phi_{i,i-1}\phi_{j,j-1} + \psi_{i-1,j-1} + \psi_{i,j} - \psi_{i,j-1} - \psi_{j,i-1} \tag{6}$$

$$\phi_{i,j} = x_i - x_j \tag{7}$$

$$\psi_{i,j} = \frac{1}{2}\phi_{i,j}^2 \ln|\phi_{i,j}| \tag{8}$$

The misfit energy term (Eq. 4) involves a summation of GSFE or Gamma energy at each nodal point. First introduced by Vitek [27], the GSFE represents the energy cost per unit area for a stacking fault which is generated by shifting the upper half-crystal relative to the lower half by a certain displacement. The GSFE can be computed using DFT [14,24]. The SVPN model provides an expedient scheme to estimate dislocation core structure and Peierls stress [24] based on DFT-determined GSFE. The GSFE represents fundamental material parameters that underlie dislocation core structure and mobility. However, there is growing evidence that in non-cubic materials, the spreading of various dislocation cores observed here could possibly indicate non-planar core structures and will result in non-Schmid behavior playing a role during dislocation glide. Thus, further modeling in this direction is necessary to accurately describe complex deformation mechanics in β-Sn. Nevertheless, since there are more than 32 different slip systems in β-Sn, it is impractical to explore all of them using first-principles atomistic calculations. Hence, our work, based on SVPN calculations, provides a much needed assessment of their likelihood in the contribution to plastic deformation.

## 3. Results and discussion

The relaxed GSFE curves for the {100}, {110} and {101} slip planes in β-Sn are shown in Figure 1 (for further detail refer to the supplemental document). Table 1 illustrates the variation in the unstable stacking fault energy ($\gamma_{usf}$) and the intrinsic stacking fault ($\gamma_{isf}$) values for the different slip systems. This anisotropic variation in the stable and unstable fault energies along with elastic properties reflects the dislocation core structure and its mobility. However, as suggested by Swygenhoven et al. [28], the absolute value of a stable stacking fault is not enough to explain the deformation mechanism; hence, the entire picture can be understood with the ratio of stable and unstable stacking fault energies. For example, when the ratio is close to unity the energy barrier for a trailing partial is low with a possibility of observing very narrow stacking width or a full dislocation as in the case of aluminum (Al). On the other hand, when the ratio is low, the energy barrier for a trailing partial is substantially higher resulting in wider stacking width as observed in copper (Cu) and nickel (Ni). Furthermore, partial dislocations separated by a narrow stacking fault width can recombine into a perfect dislocation and cross slip; whereas, a wider fault region makes cross slip energetically less favorable. For example, the earlier work of Chu and Li [29] regarding a dislocation line in β-Sn can be understood in terms of stacking fault ratio. For the slip lines on the {100} surface, Chu and Li have observed cross slip of the {110} <111> and {110} <001> slip systems at room temperature; whereas, no cross slip of the {110} <110> system was observed on the {001} surface since the stacking fault ratio was close to 0.49.

In this work, by measuring the ratio between the stable and unstable stacking fault energies, we found that the {100} <010> slip system had the widest stacking fault region followed by the {110} <$\bar{1}$10>, {101} <010> and {101} <$\bar{1}$01> slip systems (ref. Table 1). For the {101} <$\bar{1}$01> slip system, we observed dislocation dissociation into three partials with metastable states (Fig. 1b), i.e., a full dislocation along the <11$\bar{1}$> direction dissociates into two partials along the <10$\bar{1}$> and <010> directions, respectively. Then, a dislocation partial along the <010> direction further dissociates into the <01$\bar{1}$> and <011> to minimize the energy of the dislocation, i.e.,

$$\frac{1}{2}[11\bar{1}] \rightarrow \frac{1}{2}[10\bar{1}] + \frac{1}{2}[010]$$

$$\frac{1}{2}[11\bar{1}] \rightarrow \frac{1}{2}[10\bar{1}] + \frac{1}{4}[01\bar{1}] + \frac{1}{4}[011]$$

$$\frac{2a^2 + c^2}{4} > \frac{3a^2 + 3c^2}{8} \tag{9}$$

As a consequence of these non-collinear dissociation reactions, the resulting shear resistance for slip as calculated by Kinoshita et al. [2] should also decrease since they were computed with an assumption of a perfect dislocation. In general, the critical resolved shear stress for a partial is significantly different than a perfect dislocation. Further, the earlier experimental work in β-Sn [29] has shown that the full dislocations on {110} <111> and {110} <001> slip systems are energetically more favorable to cross slip while energetically, it is harder to cross slip due to dislocation dissociation on the {110} <110> slip system. Hence, these calculations reveal various dissociation reactions which are crucial to a comprehensive understanding of deformation behaviors in β-Sn. Overall, the lowest $\boldsymbol{\gamma_{usf}}$ is observed for the {101} <10$\bar{1}$>, i.e., this is the most easily activated slip system as compared to other slip systems. However, this slip system has a fractional dislocation structure with no well-defined stacking fault (local minimum) in between [30].

Next, the dislocation core properties of four dislocations, i.e., 0º (screw), 30º, 60º and 90º (edge) dislocations on the {100} slip plane with a <c> type (3.213 Å) dislocation, the {101} plane with a 1/2<2a+c> type (4.489 Å) dislocation and the {110} plane with a 1/2<2a+c> type (4.489 Å) dislocation are discussed. The disregistry vector, $\boldsymbol{f} = f_1\hat{x} + f_2\hat{y} + f_3\hat{z}$, a function of the angle $\theta$ between the dislocation line and the Burger's vector $\boldsymbol{b}$, was computed using the GSFE curve (Figure 1) along the various slip systems. Hence, it is imperative to analyze the two components of $\boldsymbol{f}$ which are parallel and perpendicular to the Burger's vector $\boldsymbol{b}$, i.e., $\boldsymbol{f}_\parallel = f_1 sin\theta + f_2 cos\theta$ and $\boldsymbol{f}_\perp = f_1 cos\theta - f_2 sin\theta$, respectively. The result of the disregistry vector and dislocation density for the screw, 30º, 60º and edge dislocations on the {100} plane are shown in Figure 2. Figures 2a and 2b show that the disregistry profile for a screw dislocation is generally much steeper as compared to the other dislocations (30º, 60º and edge).

The dislocation half core width, $\zeta$, defined as half of the atomistic distance over which $\boldsymbol{f}_\parallel$ changes from $\frac{1}{4}\boldsymbol{b}$ to $\frac{3}{4}\boldsymbol{b}$, can be obtained from Figure 2. These values of dislocation half core width are listed in Table 2 for the four dislocations, i.e., screw, 30º and 60º mixed dislocations and edge dislocations on the {100}, {101} and {110} planes. It can be observed that the $\zeta$ increases as the angle ($\theta$) between the dislocation line and Burger's vector increases. Further, by analyzing the core structure properties obtained from the different GSFE curves in Table 2, we noticed a widely spread dislocation core for dislocations on the {110} plane as compared to dislocations on the {100} and {101} planes. Furthermore, from the density plot (Figures 2c and 2d), we see a double peak for the 0˚ (screw) and 30˚ dislocations which suggest the dissociation of dislocations; whereas, for the 60˚ and 90˚ (edge) dislocations, there is only one peak suggesting no dissociation of dislocations. The narrowing or widening of the core will significantly affect the mobility of dislocations as the Peierls stress is exponentially related to the dislocation core width. Note that a widely spread core structure could also indicate a non-planar core structure.

The core energies ($U_{core}$) along with the elastic ($U_{elastic}$) and misfit ($U_{misfit}$) energies are illustrated in Figure 3 as a function of dislocation orientations ($\theta$) (or dislocation character) on the {100}, {101} and {110} planes. Here, the misfit energy increases as the angle between the dislocation line and the Burger's vector increases. This increase in the misfit has been attributed to the change in the dislocation character [19]. On the other hand, the elastic and core energies decrease with increasing angle between the dislocation line and the Burger's vector. Further, the results show that the elastic energy contribution to the overall core energy is significant and is also more sensitive to the angle than the misfit energy. Note that in some cases the elastic energy has been found to be positive, as in the case of Si [31]. Nevertheless, a similar negative energy trend has been observed in other metallic systems such as Al [19].

The results for the Peierls stress for the screw, 30º and 60º mixed dislocations and edge dislocations on the {100}, {101} and {110} planes are listed in Table 2. The Peierls stress for the various types of dislocation character on the {101} plane was higher than that for the dislocations on the other planes (i.e., {110} and {100}). In general, the Peierls stress for the screw dislocation was found to be orders of magnitude higher than the edge dislocation, i.e., the greater the edge component of a mixed dislocation, the greater the dislocation mobility (lower Peierls stress). The largest Peierls stress observed was 365 MPa for the dislocation on the {101} plane, which is in good agreement with the experimental observation, see [32]. Further, the Peierls stress for {100} and {110} planes were found to be 192 MPa and 295 MPa, respectively. Note that the Peierls stress was calculated at 0 K, which represents the lower limit of the Peierls stress in experiments at the same temperature. Overall, the narrowing of the core width, as listed in Table 2, is consistent with the increase of the Peierls stress since the Peierls stress depends exponentially on the ratio between the dislocation core width and the atomic spacing along the dislocation line.

Next, we analyze the relationship between the Peierls stress and the dislocation character. For this, we compute the Peierls stress of 20 different dislocations with the same Burger's vector but with different dislocation angles. It is possible to fit the calculated Peierls stress for 20 different dislocations to the core width using the expression derived based on the continuum elastic theory, see [33]. In general, the continuum elastic theory suggests that the Peierls stress ($\sigma_p$) can be expressed as a function of the core width, $\zeta$, and the average nodal spacing along the X direction, $\bar{a}$, by the following equation [31,34]:

$$\sigma_p = C_1 \frac{Kb}{\bar{a}} \exp(C_2 \frac{\zeta}{\bar{a}}) \tag{10}$$

$C_1$ and $C_2$ are material constants that depend on the crystal structure (such as FCC versus BCT), dislocation glide plane, etc. Figure 4 shows the calculated Peierls stress $(\ln(\sigma_p \bar{a}/Kb))$ as a function of normalized core width ($\zeta/\bar{a}$) for the various possible dislocations (the Burger's vector directions and the dislocation characters) along the {100}, {101} and {110} planes and the corresponding fitted (solid line) curves using Eq. 10. The fitted material parameters, $C_1$ and $C_2$, for different planes are listed in Table 3. In general, the continuum expression (Eq. 10) for the Peierls stress is comparable to the calculated values, except we observed a large deviation for some points, such as dislocations with an angle of 10.9° and 14.9° on the {110} plane. This type of deviation stems from the sensitivity of the Peierls stress to the average atomic spacing (non-linearity) and the core width. Similar deviations have been observed for Al [19]. Nevertheless, our results show a simple relationship between the Peierls stress and the normalized core width for the different glide planes in β-Sn. Also, for all the glide planes in the β-Sn examined here, the Peierls stress decreases monotonically as the normalized core width increases. The results further show that the slip occurs preferentially on the {100} plane as compared to the {101} and {110} planes with the {100} <001> slip system being the system with easiest slip.

Our main observations about the slip systems in β-Sn can be summarized as follows. The theoretical Schmid factor analysis (see supplemental documents) is a function of the loading direction and predicts {121} <1$\bar{1}$1> to have the highest likelihood of having high resolved shear stresses, followed by the {100} <001> system. However, according to the GSFE curves, the most easily activated slip system is {100} <001>. As a result, the relative areas of these slip systems should expand the range of orientations of likely activity depicted in Supplemental Fig. S1, and the relative area of {121} <1$\bar{1}$1> should contract. Furthermore, the Peierls stress analysis predicts slip on the {100} plane followed by {101} and then on the {110} planes with the {100} <001> slip system being the system with the easiest slip, which is in good agreement with the orientation imaging microscopy experimental work of Telang and Bieler [35]. Overall, these results provide new insights into the anisotropic plastic deformation behavior of β-Sn.

## 4. Conclusions

The dislocation mediated plasticity in β-Sn is far more than that found in cubic materials. Several experimental studies have provided evidence of various slip systems being active [9,12,36–41]. However, the dislocation core structure, separation of partials and nature of the dissociation reactions in β-Sn have rarely been studied. In fact, there is no systematic modeling or experimental study which critically assesses different non-equivalent slip systems and subsequent core properties in β-Sn. Thus, we present in this study the first modeling of dislocation cores in β-Sn using the SVPN model. In this paper, we reveal the nature of slip behavior, core structures, dislocation dissociation reactions and mobility in β-Sn. The following conclusions can be drawn from this work:

1. For the {101} <$\bar{1}01$> slip system, we observed dislocation dissociation into three partials with metastable states (Fig. 1b), i.e., a full dislocation with the Burger's vector of $\frac{1}{2}[11\bar{1}]$ into a leading partial with the Burger's vector of $\frac{1}{2}[10\bar{1}]$ and a trailing partial with the Burger's vector of $\frac{1}{2}[010]$. Furthermore, the trailing partial further dissociates into two partial dislocations, with a Burger's vector of $\frac{1}{4}[01\bar{1}]$ and $\frac{1}{4}[011]$. As a consequence of these non-collinear dissociation reactions, the resulting shear resistance for slip as calculated by Kinoshita et al. [2] should also decrease, since they were computed with an assumption of a perfect dislocation.
2. By analyzing the core structure properties obtained from the different GSFE curves in Table 2, we noticed a widely spread dislocation core for dislocations on the {110} plane as compared to dislocations on the {100} and {101} planes. Furthermore, from the density plot (Figures 2c and 2d), we see a double peak for the 0° (screw) and 30° dislocations which suggest the dissociation of dislocations. The narrowing or widening of the core will significantly affect the mobility of dislocations as the Peierls stress is exponentially related to the dislocation core width.
3. In general, the Peierls stress for the screw dislocation was found to be orders of magnitude higher than the edge dislocation, i.e., the more the edge component of a mixed dislocation, the greater the dislocation mobility (lower the Peierls stress). The largest Peierls stress observed was 365 MPa for the dislocation on the {101} plane at 0K, which is in par with the experimental observation, see [32].
4. The results further show that the slip occurs preferentially on the {100} plane followed by {101} and then on {110} planes with the {100} <001> slip system being the system with the easiest slip.

Overall, our study provides critical knowledge towards a comprehensive understanding of non-equivalent slip systems and subsequent deformation mechanisms in β-Sn. The SVPN model provides the necessary theoretical framework to utilize atomistically derived dislocation properties at the continuum scale. The simplicity and inexpensive nature of SVPN model make it easier to obtain qualitative insights on the dislocation properties across a large number of slip systems in β-Sn. However, in the future, systematic modeling efforts are necessary in the development of accurate predictive capabilities that can aid in improving the reliability of solder joints. For instance, utilizing first principle calculations to obtain a comprehensive GSFE surface along the various slip planes in β-Sn can provide accurate insights on the dislocation core properties, as shown in a recent study for hcp metals [42]. Nevertheless, in this work for the first time a systematic study was carried out to assess the dislocation properties along the various non-equivalent slip system in β-Sn.

## Acknowledgement

The authors are grateful for the financial support from the School for the Engineering of Matter, Transport, and Energy (SEMTE) at Arizona State University, the Army Research Office under (W911NF1410550) and the Office of Naval Research (N00014-15-1-2092). We also appreciate Fulton High Performance Computing at Arizona State University for enabling us to conduct our simulations.

**Table 1:** The unstable stacking fault energy ($\gamma_{usf}$), the stable stacking fault energy (i.e., the intrinsic stacking fault energy, $\gamma_{isf}$) and the ratio $\gamma_{ratio} = \gamma_{isf}/\gamma_{usf}$ for nine different slip systems in β-Sn.

| Slip system | $\gamma_{usf}$ (eV/Å²) x10⁻³ | $\gamma_{isf}$ (eV/Å²) x10⁻³ | $\gamma_{ratio}$ |
|---|---|---|---|
| {100}<011> | 12.9 | 4.3 | 0.33 |
| {100}<010> | 14.6 | 7.3 | 0.50 |
| {100}<001> | 12.4 | - | - |
| {110}<$\bar{1}$10> | 27.0 | 13.4 | 0.49 |
| {110}<001> | 13.4 | - | - |
| {110}<$\bar{1}$11> | 24.2 | - | - |
| {101}<010> | 19.0 | 11.8 | 0.62 |
| {101}<$\bar{1}$01> | 7.3 | 7.0 | 0.96 |
| {101}<11$\bar{1}$> | 14.9 | - | - |

**Table 2:** Core width and Peierls stress of four dislocations (screw, 30º, 60º and edge dislocations) on the {100}, {101} and {110} planes.

| Dislocation | Core width (Å) | | | Peierls stress (MPa) | | |
|---|---|---|---|---|---|---|
| | {100} | {101} | {110} | {100} | {101} | {110} |
| Screw | 2.5 | 1.1 | 3.2 | 192 | 365 | 295 |
| 30º | 3.1 | 2.0 | 3.8 | 45 | 50 | 40 |
| 60º | 3.3 | 2.1 | 3.9 | 80 | 140 | 71 |
| Edge | 3.8 | 2.2 | 4.1 | 32 | 38 | 33 |

**Table 3:** $C_1$ and $C_2$ constants for each plane

| Planes | {100} | {101} | {110} |
|---|---|---|---|
| $C_1$ | $2\pi$ | $\pi/2$ | $2\pi$ |
| $C_2$ | -8.6 | -5.3 | -5.1 |

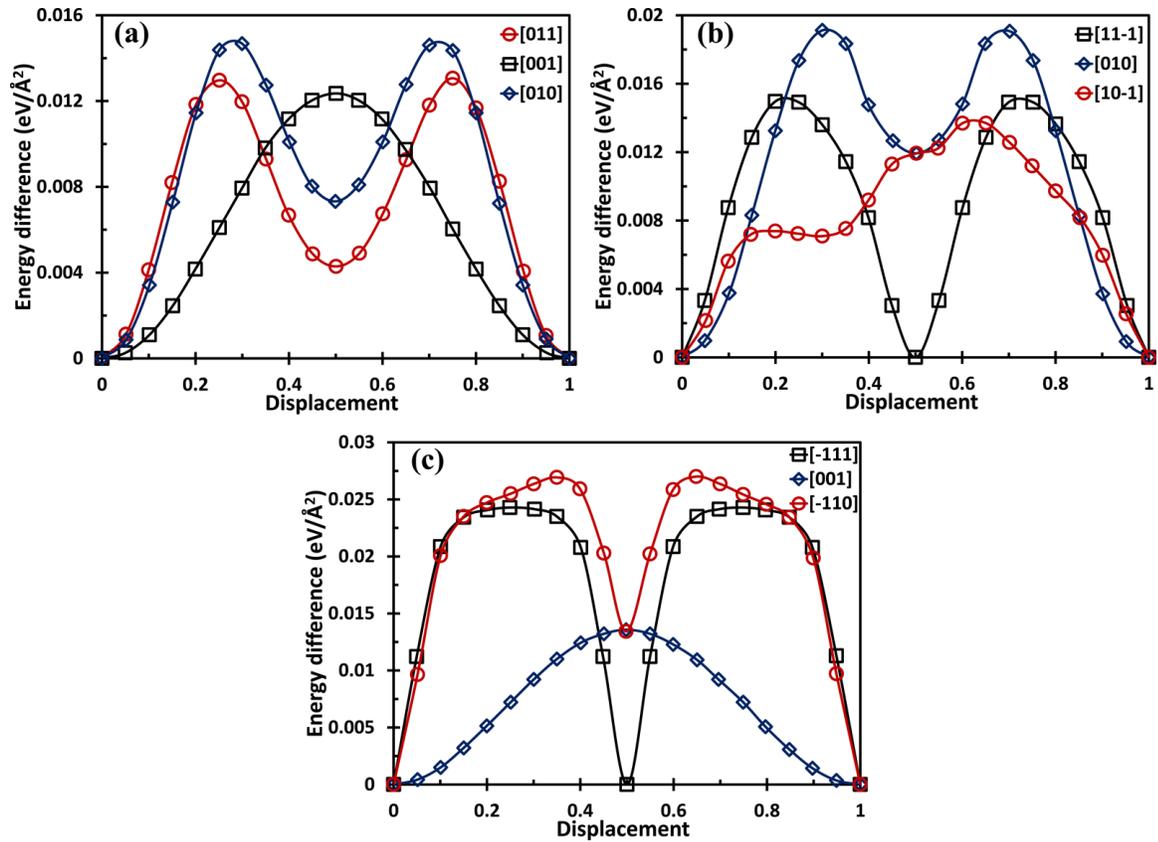

**Figure 1:** The GSF energy curves for a) (100), b) (101) and c) (110) planes in (eV/Å$^2$) obtained from DFT calculations.

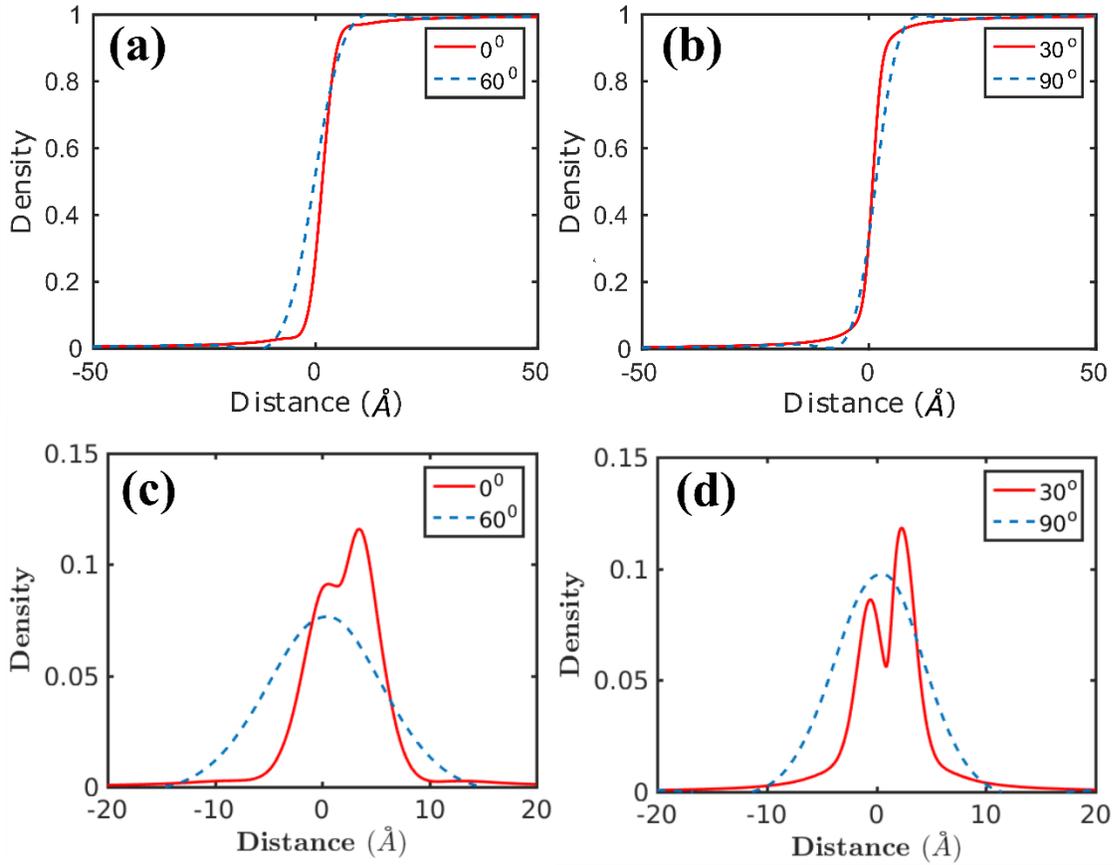

**Figure 2:** The disregistry vector and dislocation density for (a) and (c) the screw (0˚) and 60˚ dislocations, respectively, and (b) and (d) the 30˚ and edge (90˚) dislocations, respectively, on a {100} plane.

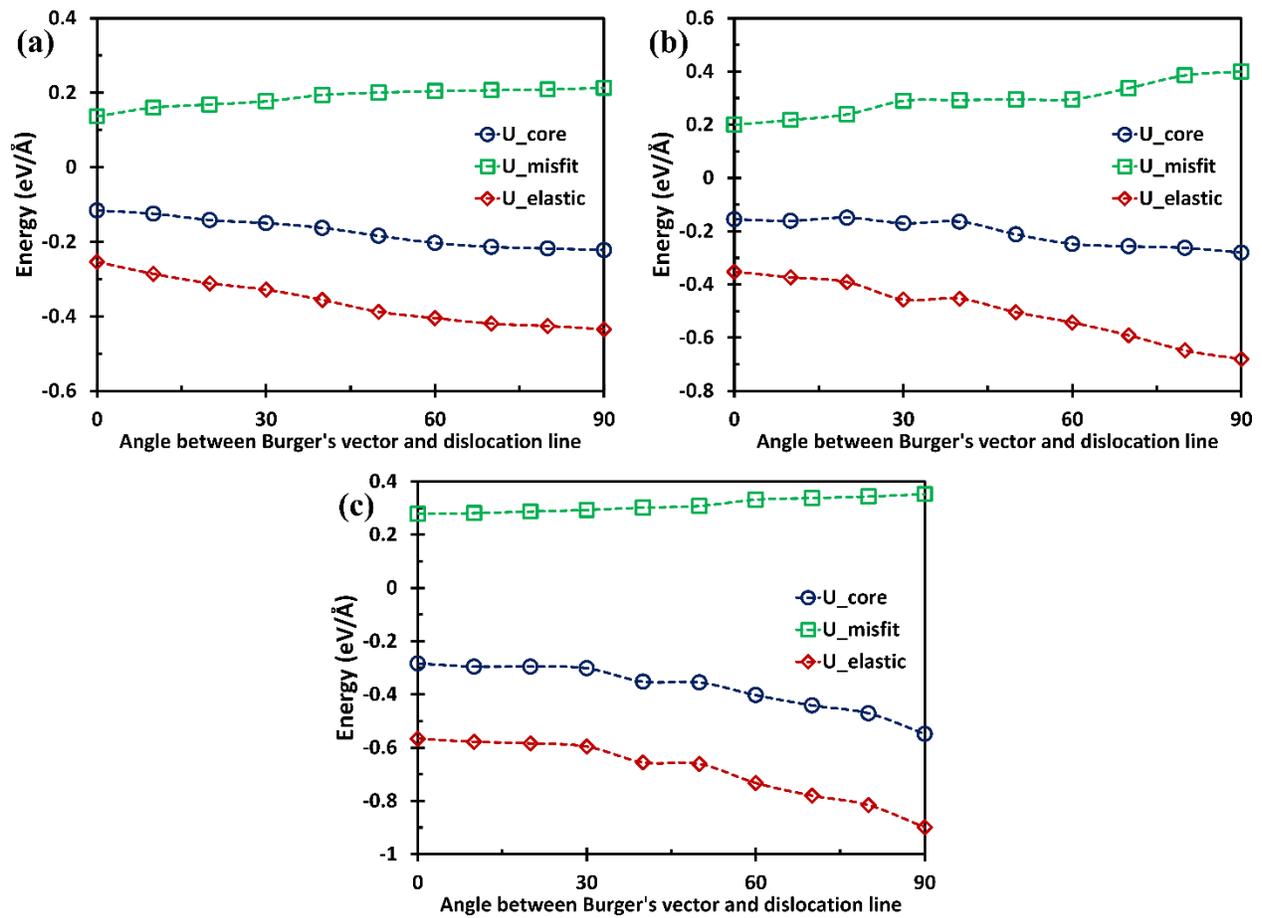

**Figure 3:** The core energy, elastic energy and misfit energy as a function of dislocation orientations along: a) the {100} plane, b) the {101} plane and c) the {110} plane.

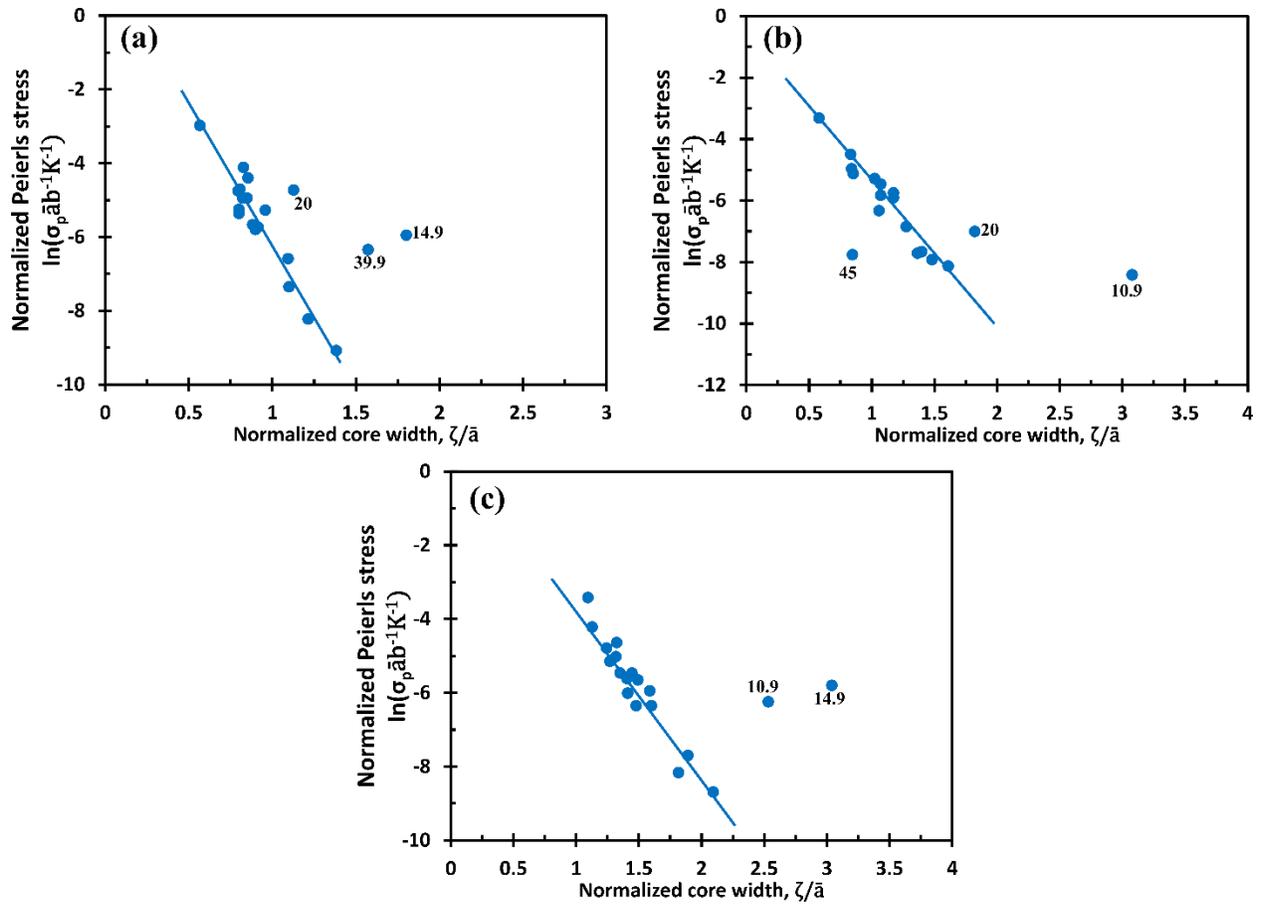

**Figure 4:** The scaled Peierls stress as a function of the ratio of the core width to the average atomic spacing perpendicular to the dislocation line along: a) the {100} plane, b) the {101} plane and c) the {110} plane.